# Relational Schema Protocol (RSP)

## Abstract

*This document specifies the Relational Schema Protocol (RSP). RSP enables loosely coupled applications to share and exchange relational data. It defines fixed message format for an arbitrary relational schema so that the changes in the data schema do not affect the message format. This prevents the interacting applications from having to be reimplemented during the data schema evolvement.*

## Status of this Memo

This Internet-Draft is submitted in full conformance with the provisions of BCP 78 and BCP 79. Internet-Drafts are working documents of the Internet Engineering Task Force (IETF). Note that other groups may also distribute working documents as Internet-Drafts. The list of current Internet-Drafts is at http://datatracker.ietf.org/drafts/current/. Internet-Drafts are draft documents valid for a maximum of six months and may be updated, replaced, or obsoleted by other documents at any time. It is inappropriate to use Internet-Drafts as reference material or to cite them other than as "work in progress." This Internet-Draft will expire on November 29, 2011.

# Table of Contents

# 1 Introduction

Widely disparate applications are very often required to share and exchange data from relational data sources. Typically, this is accomplished by passing messages over a shared environment (e.g. computer network, file system, computing memory, etc.) in a well-defined, machine-processable format. In such a case, the data schema is mapped to the message format such that the elements of the message format reflect the relations and their attributes in the data schema. If the data schema is a subject to change, the message format needs to be redefined. As a consequence, all the interacting applications need to be reimplemented. This represents a serious issue for evolving data schemas.

This specification defines a new communication protocol, RSP, for universal relational data exchange. This protocol enables the interacting applications to describe arbitrary relational schema (including the contained data) using a fixed hierarchy of data types. This way, the format of the messages does not need to be redefined when the data schema evolves.

## 1.1 Requirements Language

The key words "MUST", "MUST NOT", "REQUIRED", "SHALL", "SHALL NOT", "SHOULD", "SHOULD NOT", "RECOMMENDED", "MAY", and "OPTIONAL" in this document are to be interpreted as described in [RFC2119].

## 1.2 Protocol Overview

The RSP protocol defines three REQUIRED request-response operations (remote procedures) listed below:

- *ReadTableHeaders* - for data schema exploration
- *ReadTable* - for data and metadata retrieval
- *Submit* - for data manipulation (insertion, alteration, deletion)

The operation behaviors are specified in section 3. The message format of operation requests and responses is defined using the hierarchy of abstract data types in section 2. These abstract data types MAY be serialized in arbitrary serialization format (e.g. XML, JSON, etc.).

# 2 Data Types

This section specifies the data types used in the protocol.

## 2.1 Shared Data Types

This section specifies the data types shared by multiple operations.

### 2.1.1 TableHeader

Data type *TableHeader* represents basic metadata of particular data table. Attributes of this data type are specified in Table 1.

| Name | Definition | Data Type | Multiplicity |
|---|---|---|---|
| Description | Description of the table for the documentation purposes | string | Zero or one (OPTIONAL) |
| PluralTitle | Plural user-friendly title of the table in the specified language | string | One (REQUIRED) |
| SingularTitle | Singular user-friendly title of the table in the specified language | string | One (REQUIRED) |
| TableName | Unique name of the table within the whole data schema | string | One (REQUIRED) |

*Table 1 - Attributes of the TableHeader data type*

### 2.1.2 ArrayOfTableHeader

Data type *ArrayOfTableHeader* represents 1-dimensional list of table headers. Attributes of this data type are specified in Table 2.

| Name | Definition | Data Type | Multiplicity |
|---|---|---|---|
| TableHeader | Table header (item in the list) | TableHeader, see Table 1 | Zero or one (OPTIONAL) |

*Table 2 - Attributes of the ArrayOfTableHeader data type*

### 2.1.3 ArrayOfint

Data type *ArrayOfint* represents 1-dimensional list of integers. Attributes of this data type are specified in Table 3.

| Name | Definition | Data type | Multiplicity |
|---|---|---|---|
| int | Integer number (item in the list) | int | Zero or one (OPTIONAL) |

*Table 3 - Attributes of the ArrayOfint data type*

### 2.1.4 Field

Data type *Field* represents metadata of particular table column. Attributes of this data type are specified in Table 4.

| Name | Definition | Data type | Multiplicity |
|---|---|---|---|
| DataType | Data type of the column (int, varchar, datetime, ...) | string | One (REQUIRED) |
| Description | Description of the column for the documentation purposes | string | Zero or one (OPTIONAL) |
| ID | Unique identifier of the column within the whole data schema | string | One (REQUIRED) |
| IsAutoGenerated | True if the column is auto-generated (identity) column | boolean | One (REQUIRED) |
| IsDisplayField | True if the column from the joined table should be displayed instead of the foreign key column in the specified table | boolean | One (REQUIRED) |
| IsEditable | True if the data in the column is editable | boolean | One (REQUIRED) |
| IsForeignKey | True if the column is the foreign key | boolean | One (REQUIRED) |
| IsJoined | True if the column is a part of joined table | boolean | One (REQUIRED) |
| IsNullable | True if the column may contain null-values | boolean | One (REQUIRED) |
| IsPrimaryKey | True if the column is a part of the primary key | boolean | One (REQUIRED) |
| MaximumLength | Maximum character length (only for character data types) | long | Zero or one (OPTIONAL) |
| Name | Unique name of the column within its parent table | string | One (REQUIRED) |
| ReferencedField | Unique name of the referenced column within the referenced table (only for foreign key columns) | string | Zero or one (OPTIONAL) |
| ReferencedTable | Unique name of the referenced table (only for foreign key columns) | string | Zero or one (OPTIONAL) |
| Table | Name of the parent table | string | One (REQUIRED) |
| Title | User-friendly title of the column in the specified language | string | One (REQUIRED) |

*Table 4 - Attributes of the "Field" data type*

### 2.1.5 ArrayOfField

Data type *ArrayOfField* represents 1-dimensional list of fields. Attributes of this data type are specified in Table 5.

| Name | Definition | Data type | Multiplicity |
|---|---|---|---|
| Field | Field (item in the list) | Field, see Table 4 | Zero or one (OPTIONAL) |

*Table 5 - Attributes of the ArrayOfField data type*

### 2.1.6 ArrayOfstring

Data type *ArrayOfstring* represents 1-dimensional list of text strings. Attributes of this data type are specified in Table 6.

| Name | Definition | Data type | Multiplicity |
|---|---|---|---|
| string | Text string (item in the list) | string | Zero or more (OPTIONAL) |

*Table 6 - Attributes of the ArrayOfstring data type*

### 2.1.7 ArrayOfArrayOfstring

Data type *ArrayOfArrayOfstring* represents 2-dimensional list of text strings. Attributes of this data type are specified in Table 7.

| Name | Definition | Data type | Multiplicity |
|---|---|---|---|
| ArrayOfstring | 1-dimensional array of text | ArrayOfstring, | Zero or more (OPTIONAL) |

| | strings (item in the list) | see Table 6 | |

*Table 7 - Attributes of the ArrayOfArrayOfstring data type*

### 2.1.8 Reference

Data type *Reference* represents metadata of the referencing table. Attributes of this data type are specified in Table 8.

| Name | Definition | Data type | Multiplicity |
|---|---|---|---|
| RedField | Unique name of the referenced column within the referenced table | string | One (REQUIRED) |
| RedTable | Unique name of the referenced table | string | One (REQUIRED) |
| RingField | Unique name of the referencing column within the referencing table | string | One (REQUIRED) |
| RingFieldTitle | User-friendly title of the referencing column in the specified language | string | One (REQUIRED) |
| RingTable | Unique name of the referencing table | string | One (REQUIRED) |
| RingTablePluralTitle | User-friendly plural title of the referencing table in the specified language | string | One (REQUIRED) |

*Table 8 - Attributes of the Reference data type*

### 2.1.9 ArrayOfReference

Data type *ArrayOfReference* represents 1-dimensional list of references. Attributes of this data type are specified in Table 9.

| Name | Definition | Data type | Multiplicity |
|---|---|---|---|
| Reference | Reference (item in the list) | Reference, see Table 8 | Zero or more (OPTIONAL) |

*Table 9 - Attributes of the ArrayOfReference data type*

### 2.1.10 Table

Data type *Table* represents data and metadata from the particular data table. Attributes of this data type are specified in Table 10.

| Name | Definition | Data type | Multiplicity |
|---|---|---|---|
| Actions | List of granted permissions (actions) for the specified user and table: 1 ~ SELECT, 2 ~ INSERT, 3 ~ UPDATE, 4 ~ DELETE | ArrayOfint, see Table 3 | One (REQUIRED) |
| Fields | List of fields (columns) | ArrayofField, see Table 5 | One (REQUIRED) |
| Header | Table header | TableHeader, see Table 1 | One (REQUIRED) |
| Items | List of data items (rows) | ArrayOfArrayOfstring, | One (REQUIRED) |

| | | see Table 7 | |
|---|---|---|---|
| References | List of references (e.g. list of tables referencing specified table) | ArrayOfReference, see Table 9 | One (REQUIRED) |

*Table 10 – Attributes of the Table data type*

## 2.2 Message Format Data Types

This section specifies the data types for operation requests and responses.

### 2.2.1 ReadTableHeadersRequest

Data type *ReadTableHeadersRequest* defines the *ReadTableHeaders* operation request message format. Attributes of this data type are specified in Table 11.

| Name | Definition | Data type | Multiplicity |
|---|---|---|---|
| UserName | User name | string | One (REQUIRED) |
| Password | Password | string | One (REQUIRED) |
| Language | Preferred localization language for the data schema (ISO 639-1 two-letter code) | string | Zero or one (OPTIONAL) |

*Table 11 – Attributes of the ReadTableHeadersRequest data type*

### 2.2.2 ReadTableHeadersResponse

Data type *ReadTableHeadersResponse* defines the *ReadTableHeaders* operation response message format. Attributes of this data type are specified in Table 12.

| Name | Definition | Data type | Multiplicity |
|---|---|---|---|
| TableHeaders | 1-dimensional list of table headers | ArrayOfTableHeader, see Table 2 | One (REQUIRED) |

*Table 12 – Attributes of the ReadTableHeadersResponse data type*

### 2.2.3 ReadTableRequest

Data type *ReadTableRequest* defines the *ReadTable* operation request message format. Attributes of this data type are specified in Table 13.

| Name | Definition | Data type | Multiplicity |
|---|---|---|---|
| UserName | User name | string | One (REQUIRED) |
| Password | Password | string | One (REQUIRED) |
| TableName | Unique name of the data table | string | One (REQUIRED) |
| Language | Preferred localization language for the data schema (ISO 639-1 two-letter code) | string | Zero or one (OPTIONAL) |
| Skip | Number of data items (rows) to skip | long | One (REQUIRED) |
| Take | Number of data items (rows) to take (if 0 is specified, all found items are returned) | long | One (REQUIRED) |
| OrderExpression | Defines how the data items should be sorted (ordered) | string | Zero or one (OPTIONAL) |
| FilterExpression | Defines how the data items should be filtered | string | Zero or one (OPTIONAL) |

*Table 13 - Attributes of the ReadTableRequest data type*

### 2.2.4 ReadTableResponse

Data type *ReadTableResponse* defines the *ReadTable* operation response message format. Attributes of this data type are specified in Table 14.

| Name | Definition | Data type | Multiplicity |
|---|---|---|---|
| Table | Data and metadata from the specified data table | Table, see Table 10 | One (REQUIRED) |

*Table 14 - Attributes of the ReadTableResponse data type*

### 2.2.5 SubmitRequest

Data type *SubmitRequest* defines the *Submit* operation request message format. Attributes of this data type are specified in Table 15.

| Name | Definition | Data type | Multiplicity |
|---|---|---|---|
| UserName | User name | string | One (REQUIRED) |
| Password | Password | string | One (REQUIRED) |
| TableName | Unique name of the data table | string | One (REQUIRED) |
| Operation | Type of submit operation (1 ~ INSERT, 2 ~ UPDATE, 3 ~ DELETE) | int | One (REQUIRED) |
| Fields | List of fields | ArrayOfField, see Table 5 | One (REQUIRED) |
| Data | Data item (table row) to submit. The order of the data values within the item must match the order of appropriate fields in Fields attribute. | ArrayOfstring, see Table 6 | One (REQUIRED) |

*Table 15 – Attributes of the SubmitRequest data type*

### 2.2.6 SubmitResponse

Data type *SubmitResponse* defines the *Submit* operation response message format. Attributes of this data type are specified in Table 16.

| Name | Definition | Data type | Multiplicity |
|---|---|---|---|
| Identity | Identity of the new data item (only for INSERT operation and tables with autogenerated identity field) | String | Zero or one (OPTIONAL) |

*Table 16 - Attributes of the SubmitResponse data type*

# 3 Operations

This section specifies the operation behaviors.

## 3.1 ReadTableHeaders

Operation *ReadTableHeaders* enumerates tables accessible for the specified user and returns their headers in the specified localization language. The operation request is of type *ReadTableHeadersRequest* (see Table 11). The operation response is of type *ReadTableHeadersResponse* (see Table 12).

## 3.2 ReadTable

Operation *ReadTable* retrieves actions (access rights for specified user), header, fields (columns), items (rows) and references of the specified table. It automatically joins all the tables referenced by foreign keys from the specified table. The operation request is of type *ReadTableRequest* (see Table 13). The operation response is of type *ReadTableResponse* (see Table 14).

## 3.3 Submit

Operation *Submit* inserts, updates or deletes a single data item (table row) on the side of the operation provider. The operation request is of type *SubmitRequest* (see Table 15). The operation response is of type *SubmitResponse* (see Table 16).

# 4 IANA Considerations

This memo includes no request to IANA.

# 5 Security Considerations

## 5.1 Communication Security

The communication security in the terms of [RFC 3552] is completely a matter of the communication environment (e.g. a transfer protocol). In order to ensure confidentiality across unsecured communication environment, the RSP messages SHOULD be encrypted. Furthermore, in unreliable communication environment, the data integrity SHOULD be verified. For these reasons, the HTTPS protocol is strongly RECOMMENDED as a transfer protocol for the RSP messages in the Internet environment.

## 5.2 System Security

The RSP protocol does not ensure full system security as specified in [RFC 3552]. It only provides a user access control in order to prevent the applications from unauthorized usage: in each operation request, the user credentials (user name and password or password hash) have to be provided by the operation requester.

The major threat regarding the system security represents inappropriate usage of the RSP-based applications, namely the SQL injection. Therefore, all the input data SHOULD be filtered on the side of the operation provider for the illegal characters and expressions depending upon the selected database engine.

# 6   Normative References

[RFC2119]   Bradner, S., "Key words for use in RFCs to Indicate Requirement Levels", BCP 14, RFC 2119, March 1997.

[RFC3552]   Rescorla, E. and B. Korver, "Guidelines for Writing RFC Text on Security Considerations", BCP 72, RFC 3552, July 2003.

# Author's Address

Vojtěch Přehnal (editor)

Faculty of Informatics, Masaryk University
Botanicka 68a
Brno  602 00
Czech Republic

Email: xprehn@mail.muni.cz